# Acoustic orbital Hall effect and orbital pumping in light-metal-ferromagnet bilayers

Mingxing Wu[1,*], Shilei Ding[1], Hiroki Matsumoto[1,2], and Pietro Gambardella[1,†]

[1]*Department of Materials, ETH Zurich, 8093 Zurich, Switzerland*
[2]*Institute for Chemical Research, Kyoto University, 6110011 Uji, Japan*

Orbital currents provide a new degree of freedom for controlling magnetism, yet their interaction with lattice dynamics remains largely unexplored. Here we report a systematic investigation of the acoustic orbital Hall effect in light metals such as Ti and Cr, where surface acoustic waves generate orbital currents through phonon–orbital coupling. The acoustic orbital current in Ti exhibits higher efficiency and longer diffusion length compared to the acoustic spin current in Pt. The sign and magnitude of the rectified acoustic voltages in nonmagnetic (Ti, Cr)/ferromagnetic (Ni, Co, $Fe_xCo_{1-x}$) bilayers are determined by the product of orbital-to-spin conversion and magnetoelastic coupling efficiencies of the ferromagnet. Additionally, we find evidence for acoustic orbital pumping, whereby the excitation of ferromagnetic resonance by surface acoustic waves injects an orbital current from the ferromagnet into the nonmagnet. These results establish lattice dynamics as an efficient driver of orbital transport, opening opportunities for low-dissipation orbitronic devices that harness and sense phonons.

## I. INTRODUCTION.

The spin Hall effect (SHE) [1] and orbital Hall effect (OHE) [2–5] refer to the conversion of a charge current into spin and orbital currents, respectively. These angular momentum currents are extensively used to control magnetism in electronic devices [6–10]. Theory predicts that the OHE originates from the orbital texture of the electronic bands of crystalline materials without explicitly requiring spin-orbit coupling (SOC), whereas the SHE emerges as a byproduct of OHE and SOC [3,4,11]. Although these two effects are intertwined in the presence of SOC [12], the OHE is believed to be more widespread than the SHE, with intrinsic orbital Hall conductivities often exceeding their spin counterparts [13]. Recently, electric-field-induced orbital accumulation has been directly observed in light metals with weak SOC, such as Ti [14], V [15], Cr [16], and Mn [17]. Furthermore, orbital-to-spin conversion in layered heterostructures comprising a light-metal orbital source and a magnetic layer with finite SOC has been shown to induce strong orbital torques on the magnetization [12,18–23]. Advances in harnessing orbital currents thus provide an opportunity for developing functional orbitronic devices that do not rely on heavy metals [8,9,24,25] or combine orbital and spin effects for efficient manipulation of the magnetization [8,12,26,27].

Generating a transverse orbital current via the OHE typically requires a longitudinal electric current, which drives orbital-angular-momentum–dependent deflection of electrons through interband coherence or impurity scattering. However, recent studies have shown that alternative mechanisms, such as orbital pumping, can generate orbital currents by transferring orbital angular momentum from a magnetic layer into a material with strong orbital-to-charge conversion. This transfer can be triggered by ultrafast laser-induced demagnetization [28–30] or excitation of ferromagnetic resonance (FMR) [31–34], in analogy to spin pumping [35]. On the other hand, orbital and spin degrees of freedom differ in their coupling to the lattice: orbitals couple strongly to the crystal field, unlike spins. This strong coupling enables phenomena such as transient multiferroicity and magnetization switching driven by THz pumping of circularly polarized phonons [36,37] as well as mixing of orbital and phonon excitations [38–40]. Moreover, phonon–orbital coupling is predicted to generate strong orbital currents via lattice dynamics [41–43]. This mechanism can be revealed through the acoustic OHE [44], in analogy to the acoustic SHE recently demonstrated in a heavy-metal-ferromagnet bilayer using surface acoustic waves (SAWs) [45].

Here, we report on a systematic investigation of the acoustic OHE in different light-metal/ferromagnetic bilayers. Our study confirms the recent findings of Ref. [44] concerning the observation of the acoustic OHE in Ti/Ni and its dependence on excitation power, frequency, and magnetic field. Additionally, we show that the acoustic OHE varies significantly with the choice of nonmagnetic/ferromagnetic (NM/FM) layers. Its magnitude and sign are determined by the product of orbital-to-spin conversion and magnetoelastic coupling efficiencies of the ferromagnet. By studying different SAW propagation directions relative to the substrate cut, we find that horizontal-shear strain plays an important role in phonon–orbital coupling and, consequently, in the generation of acoustic orbital currents. Comparative measurements of Ti/Ni and Pt/Ni bilayers as a function of Ti and Pt thickness show that the acoustic OHE in Ti/Ni is characterized by a longer diffusion length in the nonmagnet and a higher efficiency than that of the acoustic SHE in Pt/Ni. Finally, we report evidence for acoustic orbital pumping from Ni into Ti, whereby an orbital current generated by magnetoelastic excitations in Ni is converted into a charge current via the inverse OHE in Ti. Based on these results, we present a phenomenological model of the acoustic OHE from the perspective of phonon–orbital coupling that accounts for the symmetry of the observed effects and provides a framework for understanding orbital effects induced by lattice dynamics.

---

[*] mingxing.wu@mat.ethz.ch
[†] pietro.gambardella@mat.ethz.ch

This paper is organized as follows: Section II describes the film growth and experimental methods; Section III presents the systematic investigation of the acoustic OHE, focusing on its dependence on layer thickness, various NM/FM combinations, and SAW propagation direction; Section IV reports the acoustic orbital pumping effect. Finally, Section V discusses the mechanism of the acoustic OHE and presents a phenomenological model of this effect based on phonon–orbital coupling.

## II. METHODS

### A. Sample preparation

Interdigital transducers (IDTs) for SAW excitation and detection were fabricated on 128° *Y*-cut LiNbO$_3$ substrates by electron beam evaporation of Ti(5)/Au(100) films (thickness in nanometers) followed by standard photolithography and lift-off processes. Each IDT has 20 pairs of fingers and 40 stripes as Bragg reflectors. The width of each finger and the gap between two fingers are 2 μm. The designed wavelength of SAWs is thus 8 μm. The distance between IDT1 and IDT2 is 220 μm. Subsequently, we deposited and patterned NM/FM bilayers (NM = Ti, Cr, Pt; FM = Ni, Co, Co$_x$Fe$_{1-x}$) in between two IDTs using DC magnetron sputtering and lift-off. An oxygen plasma was used to clean the substrate before film deposition. The sputtering Ar pressure was set to 2.6 mTorr, and the power ranged from 7 to 14 W depending on the target material. The NM/FM bilayers were protected against oxidation by an 8-nm-thick Si$_3$N$_4$ capping layer deposited in situ with an rf sputter gun.

### B. Generation and detection of acoustic orbital and spin currents

Figures 1(a) and (b) present schematic illustrations for the generation and detection of the acoustic orbital and spin currents. NM/FM bilayers are patterned in the shape of Hall bars between two IDTs. An ac electric field applied to IDT1 generates SAWs that propagate coherently along the LiNbO$_3$ surface. The interaction between lattice phonons in the SAW and orbital (spin) degrees of freedom in the nonmagnetic layer generates an ac orbital (spin) current that diffuses towards the ferromagnetic layer, where it is either absorbed or reflected depending on whether the orbital (spin) polarization is oriented transverse or parallel to the magnetization. As the SAW also modulates the orientation of the magnetization via magnetoelastic coupling, the reflected orbital (spin) current has a rectified dc component that results in charge accumulation via the inverse OHE (SHE), which can be detected by measuring the SAW-induced longitudinal voltage $V_{xx}$. This generation and detection scheme is analogous to that employed to measure the acoustic SHE in Pt/CoFeB bilayers [45].

We characterized the SAW propagation using a vector network analyzer to measure the transmission spectrum $S_{21}$ from IDT1 to IDT2, as shown in Fig. 1(c). The peak frequency emerges at ~463 MHz, corresponding to a SAW wavelength of 8 μm. A time domain gating function was applied to exclude signal contributions due to electromagnetic waves propagating through air. A radiofrequency (rf) signal with a power of 20 dBm was applied to IDT1 for exciting SAWs at their resonance frequency. The rf signal was modulated with a frequency of 2111 Hz. We define the *x*-axis as parallel to the SAW wavevector *k*, with the magnetic field used to orient the magnetization rotating in the film plane at an angle $\varphi$ relative to the *x*-axis. The longitudinal voltage was then detected by a lock-in amplifier referenced to the rf signal modulation. For the acoustic OHE measurements, we rotated the devices in a magnetic field of 30 mT. In the acoustic orbital pumping measurements, the magnetic field was swept at different angles. Each measurement was averaged several times to improve the signal-to-noise ratio. All the measurements were performed at room temperature.

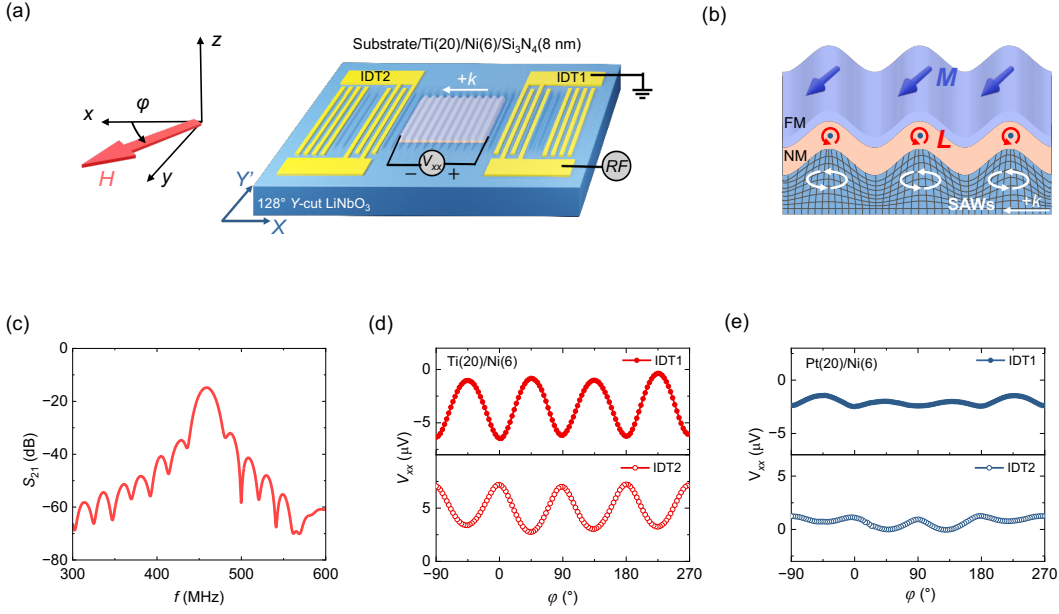

FIG. 1. (a) Setup used for acoustic OHE and orbital pumping measurements. (b) Schematic of SAW-driven acoustic OHE. $L$ and $M$ denote the orbital angular momentum and magnetization, respectively. FM and NM denote the ferromagnetic and nonmagnetic layers. The white circular arrows depict rotational motion of the lattice induced by SAWs. (c) Transmission spectrum $S_{21}$ of SAWs with $k \parallel X$. (d) Representative measurement of the acoustic voltage in a Ti(20)/Ni(6) device with $k \parallel X$ as a function of magnetic field angle. IDT1 (IDT2) denotes that the rf signal is applied to IDT1 (IDT2). (e) Measurement of the acoustic SHE in a Pt(20)/Ni(6) device with $k \parallel X$. The input rf power was set to 20 dBm for SAW excitations and the amplitude of the in-plane magnetic field was set to 30 mT.

## III. ACOUSTIC ORBITAL HALL EFFECT

### A. Longitudinal voltage due to the acoustic OHE and SHE

To probe the acoustic OHE, we measure the variation of $V_{xx}$ as a function of $\varphi$. We focus first on measurements performed on Ti/Ni bilayers as a model system for studying OHE-related effects. Figure 1(d) shows a representative measurement of $V_{xx}$ in a Ti(20)/Ni(6) device. The applied magnetic field is set to 30 mT to suppress contributions from acoustic orbital pumping, which we discuss later. The measured voltage varies as $V_{xx} = V_{xx}^0 + V_{xx}^{\text{AOHE}} \sin^2(2\varphi)$ and changes sign upon reversal of $k$. The term $V_{xx}^0$ represents a constant rectification offset arising from the acoustoelectric effect [45,46], whereas the four-fold symmetric component with amplitude $V_{xx}^{\text{AOHE}}$ is a hallmark of the acoustic OHE, analogous to the acoustic SHE [45]. Inverting the SAW propagation direction by switching the input rf signal from IDT1 to IDT2 changes the sign of both offset and acoustic OHE voltages. Additional characterizations of the acoustic OHE are presented in Appendix A, including its dependences on the applied magnetic field, input power, and IDT frequency. Specifically, $V_{xx}^{\text{AOHE}}$ decreases with increasing magnetic field, reflecting the suppression of the SAW-induced oscillations of the magnetization in Ni. Moreover, it scales with the input power and the resonance frequency of IDTs. These properties are analogous to those reported for the acoustic SHE in Pt/CoFeB [45] and acoustic OHE in Ti/Ni bilayers [44].

By comparing measurements performed on Ti(20)/Ni(6) and Pt(20)/Ni(6) devices representative of the acoustic OHE and SHE, respectively, we show that these two effects differ in several key aspects. Figure 1(e) presents the acoustic voltage measured in a Pt(20)/Ni(6) device. The amplitude of the acoustic voltage induced by the SHE in Pt is only 30% of the acoustic OHE voltage in Ti(20)/Ni(6), despite Pt possessing a spin Hall conductivity 5-10 times larger than the orbital Hall conductivity of Ti [12,20,47], as confirmed by spin-orbit torque measurements (not shown). This comparison suggests that the phonon–orbital coupling in a light metal with weak SOC such as Ti is much stronger than spin–phonon coupling in heavy metals.

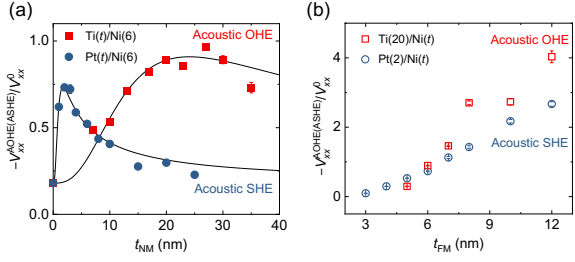

FIG. 2. Thickness dependence of the acoustic OHE and SHE. (a) $-V_{xx}^{\text{AOHE(ASHE)}}/V_{xx}^0$ ratios as a function of Ti (Pt) thickness. The black lines are fits using Eq. (1) to extract the orbital and spin diffusion lengths for Ti and Pt, respectively. (b) $-V_{xx}^{\text{AOHE(ASHE)}}/V_{xx}^0$ ratios as a function of Ni thickness for Ti/Ni and Pt/Ni bilayers. The magnetic field was set to 30 mT. The error bars are the standard deviation from the fits.

### B. Thickness dependence of the acoustic OHE and SHE

Another important distinction between the acoustic OHE and SHE emerges from the diffusion lengths of the orbital and spin currents in the nonmagnetic layer ($t_N$). To compare different devices, we normalize the acoustic OHE (SHE) voltage $V_{xx}^{\text{AOHE(ASHE)}}$ by $V_{xx}^0$, which accounts for variations in SAW transmission across samples [45]. Figure 2(a) shows $-V_{xx}^{\text{AOHE(ASHE)}}/V_{xx}^0$ as a function of $t_N$. Notably, the acoustic OHE signal continues to rise in Ti/Ni until $t_{\text{Ti}}$ exceeds 20 nm, whereas the acoustic SHE signal in Pt/Ni saturates already at $t_{\text{Pt}} = 2$ nm. The corresponding orbital (spin) diffusion length can be extracted using a drift-diffusion model [48,49]:

$$\frac{V_{xx}^{\text{AOHE(ASHE)}}}{V_{xx}^0} = -\frac{\Omega}{1+\xi}\frac{\lambda_N}{t_N}\tanh^2\left(\frac{t_N}{2\lambda_N}\right) \times \tanh\left(\frac{t_N}{\lambda_N}\right) + C_1, \quad (1)$$

where $\Omega$ is a prefactor related to the orbital (spin) Hall angle and phonon–orbital (spin) coupling of the nonmagnetic layer, $\lambda_N$ the orbital (spin) diffusion length of the nonmagnetic layer, $\xi = \frac{\rho_N t_F}{\rho_F t_N}$ the shunting current ratio, $\rho_{N,F}$ and $t_{N,F}$ the resistivity and thickness of nonmagnetic and ferromagnetic layers, respectively, and $C_1$ the offset due to the self-induced acoustic voltage from the Ni layer (Appendix B). Taking $\rho_{\text{Ti}} = 200$ μΩcm, $\rho_{\text{Ni}} = 51$ μΩcm, and $\rho_{\text{Pt}} = 31$ μΩcm from measurements, and $C_1 = 0.18$ as measured for Ni(6) single layer, we fit the data in Fig. 2(a) using Eq. (1), which gives $\lambda_{\text{Ti}} = 7.3 \pm 0.5$ nm and $\lambda_{\text{Pt}} = 0.45 \pm 0.04$ nm. The value of $\lambda_{\text{Ti}}$ is comparable to the relatively long orbital diffusion length derived from orbital Hall magnetoresistance (OMR) [50] and orbital pumping [32] in Ti/Ni, whereas $\lambda_{\text{Pt}}$ agrees with the short spin diffusion length of Pt and previous measurements of the acoustic SHE [45]. Additionally, Fig. 2(b) shows that both the acoustic OHE and SHE increase monotonically with Ni thickness, as expected due to the enhancement of magnetoelastic coupling in thicker magnetic layers [51].

### C. Acoustic OHE in various NM/FM bilayers

In general, the orbital currents generated by a nonmagnetic layer cannot directly interact with the magnetization of an adjacent ferromagnet, requiring SOC-mediated orbital-to-spin conversion in the magnetic layer [18,52,53]. Measurements of orbital torques have shown that the orbital-to-spin conversion efficiency, $c_{\text{L-S}}$, varies significantly across different magnetic elements and their alloys [8,12,19]. We thus investigated the acoustic OHE in Ti combined with different ferromagnetic layers and in Cr/Ni. As shown in Fig. 3(a), both $V_{xx}^{\text{AOHE}}$ and $V_{xx}^0$ exhibit a strong material dependence. The corresponding acoustic OHE efficiency, quantified by the ratio $-V_{xx}^{\text{AOHE}}/V_{xx}^0$, is shown in Fig. 3(b). Remarkably, this ratio changes not only in magnitude but also in sign. Such a trend cannot be solely attributed to variations in $c_{\text{L-S}}$, since reported values of $c_{\text{L-S}}$ have the same sign and differ in magnitude by at most a factor of four in the materials considered here (green bars in Fig 3(c), data from Ref. [18]). Since the acoustic voltage originates from rectification of the orbital current via SAW-induced magnetization oscillations, it is also expected to scale with the effective magnetoelastic coupling, defined as $b_{\text{ME}} = B/M_s$, where $B$ is the magnetoelastic coupling constant in units of J/m$^3$ and $M_s$ the saturation magnetization. Polycrystalline Ni and Co exhibit positive $b_{\text{ME}}$, whereas FeCo alloys typically show composition-dependent negative $b_{\text{ME}}$ (light blue bars in Fig. 3(c), data from Refs. [54–56]). Accordingly, we find that $-V_{xx}^{\text{AOHE}}/V_{xx}^0$ scales with the product $c_{\text{L-S}} \cdot b_{\text{ME}}$, as illustrated in Fig. 3(d). In addition, the acoustic OHE efficiency in Cr/Ni is smaller compared to Ti/Ni.

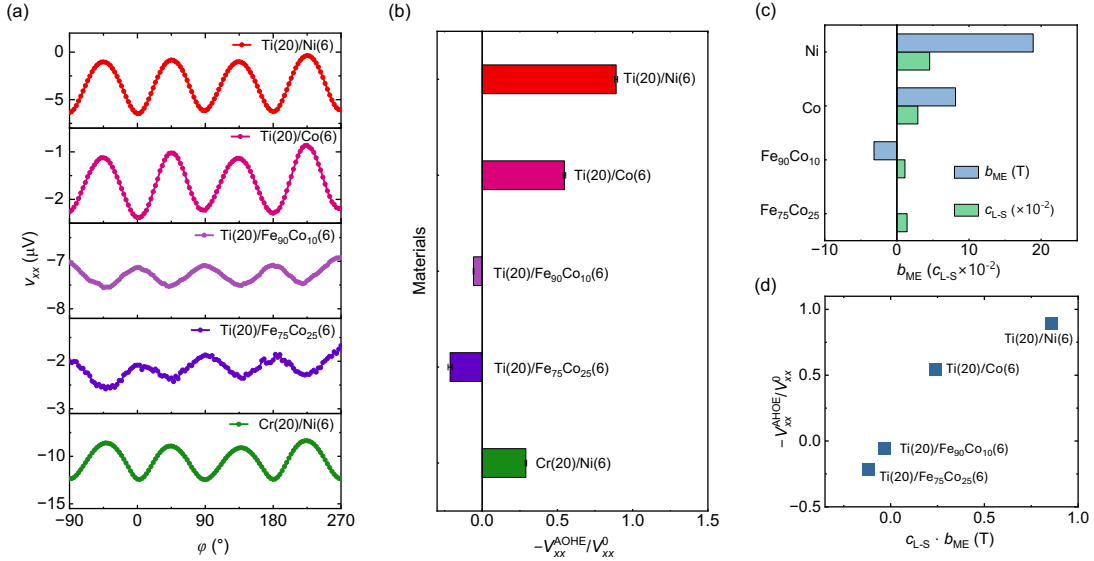

FIG. 3. Acoustic OHE in various NM/FM bilayers. (a) Variation of $V_{xx}$ and $V_{xx}^0$ in different material combinations. (b) Corresponding efficiency of the acoustic OHE as given by the ratio $-V_{xx}^{AOHE}/V_{xx}^0$. The error bars are the standard deviation from the fits. (c) Calculated orbital-spin conversion efficiency $c_{L-S}$ and magnetoelastic coefficient $b_{ME}$ for Ni, Co, $Fe_{90}Co_{10}$, and $Fe_{75}Co_{25}$. Data from Refs. [18,54–56]. (d) $-V_{xx}^{AOHE}/V_{xx}^0$ as a function of the product $c_{L-S} \cdot b_{ME}$ for the Ti-based devices. The magnetic field was set to 30 mT for these measurements.

Thus, the acoustic OHE can be maximized by selecting a nonmagnet with strong phonon–orbital coupling (such as Ti) and a ferromagnet with large $c_{L-S} \cdot b_{ME}$ product (such as Ni)[22,40].

### D. Dependence of the acoustic OHE on SAW propagation direction

To investigate the dependence of the acoustic OHE on substrate orientation, we fabricated Ti(20)/Ni(6) devices with the SAW propagation wave vector $k$ oriented at different angles $\alpha$ relative to the $X$-axis of LiNbO$_3$, as shown in Fig. 4(a). Apart from the SAW propagation direction, all devices were identical and patterned on the same chip. Figure 4(b) shows exemplary results obtained for $\alpha = 0$ ($k \parallel X$), $\alpha = 20°$, $\alpha = 40°$, and $\alpha = 90°$ ($k \parallel Y'$). Significant changes in $V_{xx}^0$ and $V_{xx}^{AOHE}$ are observed with $k$ varying from $X$ to $Y'$. Remarkably, as $\alpha$ increases from 0 to 90°, the four-fold symmetric signal shifts towards larger $\varphi$, leading to a sign reversal of $V_{xx}^{AOHE}$ for $k \parallel Y'$ relative to $k \parallel X$. To capture this feature, we introduce a phase term $\delta$ and fit the longitudinal voltages using the expression:

$$V_{xx} = V_{xx}^0 + V_{xx}^{AOHE} \sin^2(2\varphi + \delta). \quad (2)$$

Figures 4(c) and (d) present the extracted magnitudes of $V_{xx}^0$ along with $V_{xx}^{AOHE}$, and the phase shift $\delta$ as a function of $\alpha$, respectively. $V_{xx}^0$ is comparable at $\alpha = 0$ and $\alpha = 90°$ with opposite signs. This indicates a comparable longitudinal strain $\varepsilon_{xx}$ between $k \parallel X$ and $k \parallel Y'$, consistent with the simulations of SAWs in 128° $Y$-cut LiNbO$_3$ reported in Ref. [57]. On the other hand, $V_{xx}^{AOHE}$ at $\alpha = 0$ is six times larger than at $\alpha = 90°$. This behavior resembles that of phonon-magnon coupling, which is only pronounced with $k \parallel X$ due to the strong shear-horizontal strain $\varepsilon_{xy}$ in this direction [51]. On the 128° $Y$-cut LiNbO$_3$ substrate, the Rayleigh wave is dominant and $\varepsilon_{xy}$ is typically much smaller than the longitudinal strain $\varepsilon_{xx}$; however, the SAW's type is strongly influenced by the periodic grating structure (IDTs and Bragg reflectors) and propagation direction ($k$ vector) [58,59]. Simulations by Hwang et al. (Supplementary Material of Ref. [51]) showed that $\varepsilon_{xy}$ is significantly larger at $\alpha = 0$ than at $\alpha = 90°$, and can be further enhanced by Bragg reflectors, reaching values comparable to $\varepsilon_{xx}$. In this case, we infer that $\varepsilon_{xy}$ plays an important role in the acoustic orbital Hall voltage. We thus incorporated $\varepsilon_{xy}$ contribution to the model of the acoustic OHE presented in Section V, where the derived expression for $V_{xx}^{AOHE}$ also includes phase terms that account for the observed sign change as a function of $\alpha$.

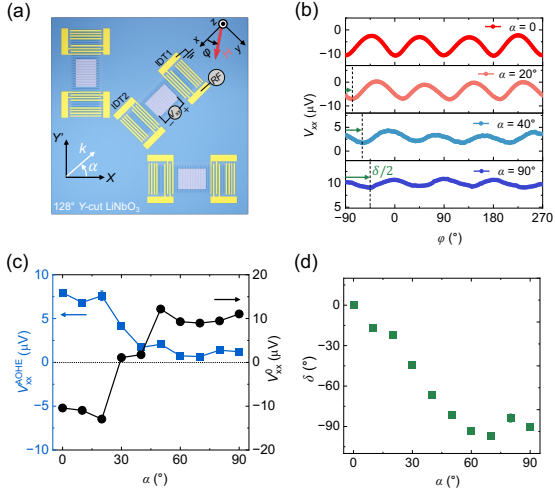

FIG. 4. Dependence of the acoustic OHE on SAW propagation direction. (a) Schematic illustration of Ti(20)/Ni(6) devices with different orientations of $k$ relative to the substrate, as defined by the angle $\alpha$. For each device, we performed measurements of $V_{xx}(\varphi)$ under a magnetic field of 30 mT. (b) Exemplary results for different values of $\alpha$. (c) Extracted $V_{xx}^0$, $V_{xx}^{AOHE}$, and phase shift $\delta$ as a function of $\alpha$. The error bars are the standard deviation from the fits.

## IV. ACOUSTIC ORBITAL PUMPING EFFECT

### A. Symmetry of acoustic OHE and orbital pumping

In addition to the acoustic OHE, other orbital effects emerge in Ti/Ni owing to the magnetoelastic excitation of the magnetization and the OHE in Ti. Orbital pumping refers to the generation of an orbital current through FMR excitation driven by a microwave electromagnetic field [31–34]. Here, we report the acoustic version of this effect, whereby the orbital pumping is induced by SAWs via the magnetoelastic coupling, analogous to the acoustic spin pumping effect due to the SHE in Pt/Ni bilayers [60]. Consistent with models of spin pumping, we find that this effect only becomes prominent as the applied magnetic field meets the FMR condition at the frequency of the SAW.

In general, both acoustic OHE and acoustic orbital pumping contribute to the longitudinal voltage $V_{xx}$ and exhibit fold-fold symmetry. Additionally, the OMR [50] can contribute an additional two-fold term due to the acoustic electric field [45]. The complete expression accounting for the longitudinal acoustic voltage in the $k \parallel X$ device thus reads:

$$V_{xx} = V_{xx}^0 + V_{xx}^{AOHE} \sin^2(2\varphi) \\
+ V_{xx}^{Pump} \sin^2(2\varphi)\sin\varphi \\
+ V_{xx}^{OMR} \cos^2(\varphi), \quad (3)$$

where $V_{xx}^{Pump}$ and $V_{xx}^{OMR}$ are the amplitudes of the orbital pumping and OMR voltage, respectively. Here, we use the devices with $k \parallel X$, so that $\delta = 0$. Figure 5(a) schematically illustrates the symmetry of each term with respect to $\varphi$. Figure 5(b) shows the angular dependence of $V_{xx}$ at magnetic fields of 6 mT and 70 mT in Ti(20)/Ni(6). The magnitude of each contribution can be extracted by fitting $V_{xx}$ with Eq. (3). The fit yields $V_{xx}^{Pump}/V_{xx}^{AOHE} = -0.3$ and $V_{xx}^{OMR}/V_{xx}^{AOHE} = 0.05$ at 6 mT, while at 70 mT, the corresponding values are -0.01 and 0.05, respectively. The pumping signal is significant at 6 mT but vanishes at 70 mT, as is only pronounced when the magnetic field satisfies the acoustic FMR condition. The OMR term is negligible and shows no dependence on the applied magnetic field, consistent with previous work [45]. Note that all the curves reported in Figs. 1-4 were measured at 30 mT and have the $\sin^2(2\varphi)$ symmetry expected of the acoustic OHE with negligible contribution of orbital pumping, as expected in off-resonance conditions.

To precisely determine the acoustic FMR condition, we measured $V_{xx}$ by sweeping the magnetic field at $\varphi = 0$ and 45°, as shown in Fig. 5(c). The acoustic voltage at $\varphi = 45°$ (black line) exhibits both an antisymmetric (blue line) and a symmetric component (red line). According to Eq. (3), the voltages due to acoustic orbital pumping and acoustic OHE are respectively odd and even under magnetization reversal. Therefore, the antisymmetric component corresponds to the acoustic orbital pumping, which is maximized at $\varphi = 45°$ [60–62]. The FMR field corresponds to the peaks of the antisymmetric $V_{xx}$ and is about 1.5 mT at $\varphi = 45°$. At $\varphi = 0$, both acoustic orbital pumping and acoustic OHE should be zero. However, when sweeping the magnetic field, we detect a symmetric signal centered around zero, which vanishes above 20 mT (gray line). We attribute this signal to the misalignment of the magnetization during the reversal process, similar to the magnetoresistance signal originating from magnetization switching.

Besides the dc detection, we confirm the acoustic FMR by measuring the absorption spectrum during the SAW's transmission ($\Delta S_{21}$). Figure 5(d) shows the $\Delta S_{21}$ in a Ti(20)/Ni(6) device at $\varphi = 0$ (gray line) and $\varphi = 45°$ (black line). At $\varphi = 0$, the transmission exhibits a dip near zero field due to the magnetization reversal process, consistent with the dc measurement in Fig. 5(c). At $\varphi = 45°$, two minima occur at $\mu_0 H \approx \pm 3$ mT, in correspondence with the FMR field expected for the Kittel mode of the Ni layer. The resonance field is comparable to, but slightly larger than that measured via the dc acoustic orbital pumping voltage, possibly due to variations in magnetic field calibration across the measurement setups or different backgrounds in the measurements. However, unlike the dc acoustic voltage, the absorption spectrum is symmetric under magnetic field inversion. This is because the transmission does not involve the inverse OHE, which has odd parity under magnetization reversal [62,63].

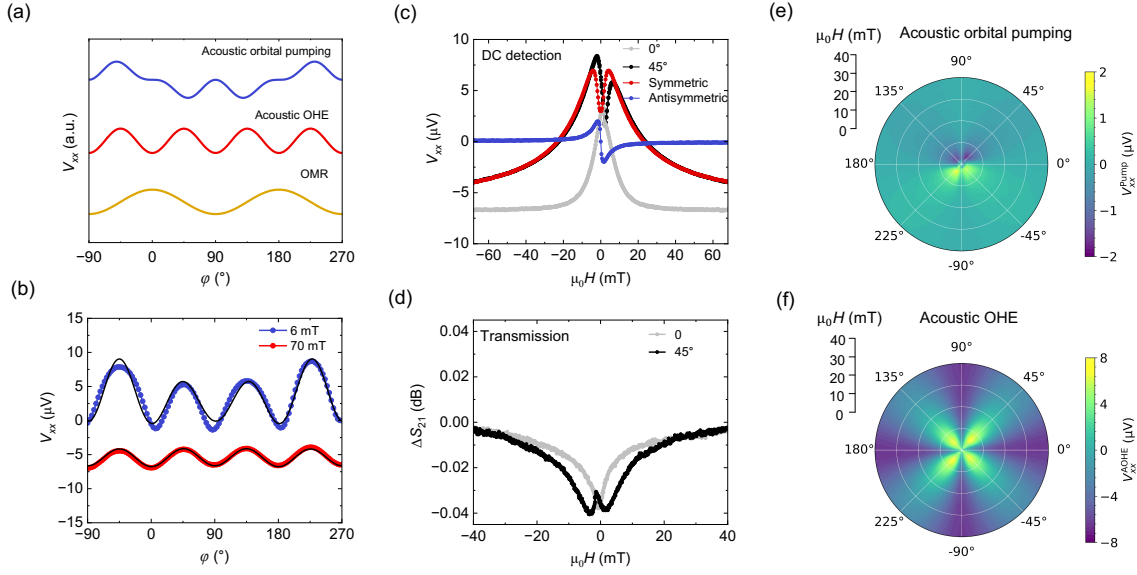

FIG. 5. Acoustic orbital pumping. (a) Calculated angular dependence of the different contributions to the acoustic voltage according to Eq. (3). The curves are arbitrarily scaled and offset for clarity. (b) Measurement of $V_{xx}$ as a function of $\varphi$ under magnetic fields of 6 mT and 70 mT in the Ti(20)/Ni(6) device. Solid lines are fits to the data using Eq. (3). The acoustic orbital pumping component is only pronounced at 6 mT near the FMR condition. (c) $V_{xx}$ as a function of magnetic field at $\varphi = 0°$ and $45°$. The curve at $\varphi = 45°$ is decomposed into symmetric and antisymmetric components corresponding to the acoustic OHE and acoustic orbital pumping, respectively. (d) SAW's absorption spectrum ($\Delta S_{21}$) as a function of magnetic field at $\varphi = 0°$ and $45°$. (e) and (f) Polar maps of the acoustic orbital pumping and acoustic OHE signals.

By sweeping the magnetic field at different in-plane angles, we obtain polar maps of the acoustic orbital pumping in Fig. 5(e) and acoustic OHE in Fig. 5(f). Clearly, both signals exhibit four-fold symmetry, with acoustic orbital pumping showing odd parity and acoustic OHE showing even parity with respect to $\varphi$. The signals reach their maximum at $\varphi = 45°$, corresponding to the strongest magnetoelastic coupling.

## B. Ti thickness dependence of orbital pumping

The magnitude of the acoustic pumping current can be defined as $I_{xx}^{\text{Pump}} = |V_{xx}^{\text{Pump}}|/R_{xx}$, where $R_{xx}$ is the resistance of the Ti/Ni bilayer. Figure 6 shows that $I_{xx}^{\text{Pump}}$ increases with increasing $t_{\text{Ti}}$, and saturates when $t_{\text{Ti}}$ exceeds 15 nm. Analogous to spin pumping in a metallic bilayer [64], $I_{xx}^{\text{Pump}}$ can be expressed as:

$$I_{xx}^{\text{Pump}} = A \frac{\lambda_N}{t_N} \tanh\left(\frac{t_N}{2\lambda_N}\right) + C_2, \quad (4)$$

where $A$ is a prefactor that depends on the orbital Hall conductivity and magnetoelastic coupling parameters (Appendix C) and $C_2 = 12$ nA is the self-induced acoustic current in a single Ni layer (Appendix B). By fitting the data in Fig. 6 using Eq. (4), we obtain $\lambda_{\text{Ti}} = 4.6 \pm 1.0$ nm, which is comparable but smaller than that extracted from the acoustic OHE. The origin of this difference is presently unknown.

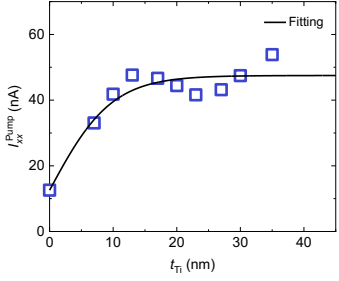

FIG. 6. Acoustic orbital pumping current as a function of Ti thickness. The black curve is a fit using Eq. (4).

## V. PHENOMENOLOGICAL MODELS OF THE ACOUSTIC OHE

### A. Phonon–orbital coupling

We discuss here different microscopic mechanisms that can drive the acoustic OHE. Recent theoretical [41–43] and experimental studies [38–40] have highlighted the strong coupling between phonons and orbital degrees of freedom through lattice dynamics. Assuming a dimensionless coefficient $\chi_{po}$ to describe the phonon–orbital coupling strength, a phenomenological expression for the orbital current induced by acoustic lattice motion can be written as (Appendix C):

$$j_{o,i}^{\zeta} = -\chi_{po} n e \epsilon_{\zeta ij} \frac{\partial u_j}{\partial t}, \quad (5)$$

where $j_{o,i}^{\zeta}$ is the orbital current density flowing along $i$ with polarization direction $\zeta$, $n$ the electron density, $e$ the electron charge, $\epsilon_{\zeta ij}$ the Levi-Civita symbol, $u_j$ the lattice displacement along $j$. Analogously to the acoustic SHE [45], the expression for the dc acoustic voltage can be derived based on the rectification of $j_{o,i}^{\zeta}$. Such a rectification occurs via magnetoresistive effects, requiring the magnetization of the adjacent magnetic layer to oscillate at the SAW's frequency, a condition consistently fulfilled by magnetoelastic coupling. Unlike the acoustic SHE, however, which relies on SOC to convert phonon-driven lattice dynamics into spin currents, the acoustic OHE directly exploits phonon–orbital coupling to efficiently excite an angular momentum current.

The phonon-mediated generation of the acoustic orbital current differs from the electrically induced OHE. The latter is primarily attributed to the intrinsic momentum-space orbital texture [3,4]. The orbital current operator in response to the external electric field is defined analogously to the spin current as $\hat{j}_i^{L\alpha} = \frac{1}{2}\{\hat{v}_i, \hat{L}_\alpha\}$, where $\hat{v}_i$ is the velocity operator and $\hat{L}_\alpha$ the orbital angular momentum operator [4,65]. However, the acoustic orbital current cannot be associated to odd-symmetric $\hat{L}_\alpha$ due to the time-reversal-even nature of lattice displacements. Instead, an even-symmetric orbital angular position operator $\{\hat{L}_\alpha, \hat{L}_\beta\}$ serves as a more appropriate quantity for capturing the phonon–orbital coupling [41,42]. The operator $\{\hat{L}_\alpha, \hat{L}_\beta\}$ measures the orbital torsion away from the equilibrium orientation of the orbital states. Such orbital torsion can be efficiently triggered by phonons, as orbital degrees of freedom are strongly coupled to the local crystal field [43]. More specifically, phonons modulate bonding lengths and thereby orbital hoppings in crystals. As a result, an alternate orbital current is induced from lattice dynamics [42]. Orbital torsion requires a combination of longitudinal and shear strains. This may explain the dependence of the acoustic OHE on the SAW propagation direction reported in Section III.

### B. Rectified voltage from the acoustic OHE

Based on Eq. (5), the longitudinal voltage from orbital current rectification can be derived using magnetoelastic coupling and drift-diffusion models [45,49]. The magnetoelastic energy reads:

$$E_{ME} = B(\varepsilon_{xx} m_x^2 + 2\varepsilon_{xy} m_x m_y + 2\varepsilon_{xz} m_x m_z), \quad (6)$$

where $\boldsymbol{m} = (m_x, m_y, m_z)$ is the unit vector of the magnetization. For simplicity, we consider an isotropic material with magnetoelastic constant $B = B_1 = B_2$, longitudinal strain $\varepsilon_{xx}$, and shear strains $\varepsilon_{xy}$, $\varepsilon_{xz}$. The detailed derivation, presented in Appendix C, gives Eq. (2) and the magnitude of the acoustic OHE voltage $V_{xx}^{AOHE} = \Gamma(1+\eta^2)^{\frac{1}{2}}[(\varepsilon_{xx}^0)^2 + (2\varepsilon_{xy}^0)^2]$, where $\Gamma$ is a pre-factor and $\eta$ is the ratio of the orbital polarization along $x$ and $y$. In addition, the phase term $\tan 2\delta = -\frac{\eta \varepsilon_{xx}^0 - 2\varepsilon_{xy}^0}{\varepsilon_{xx}^0 + 2\eta \varepsilon_{xy}^0}$ induces a shift in the four-fold symmetric signals (Fig. 4).

Obviously, $V_{xx}^{AOHE}$ is strongly influenced by $\varepsilon_{xx}^0$, $\varepsilon_{xy}^0$ and $\eta$. Therefore, we simulate $V_{xx}$ in the $k \parallel X$ and $k \parallel Y'$ geometries with different values of $\varepsilon_{xy}$ to investigate its effect on the acoustic OHE. We take the strain parameters $\varepsilon_{xx}^0 = 10^{-5}$ for $k \parallel X$ and $\varepsilon_{xx}^0 = -10^{-5}$ for $k \parallel Y'$; $\varepsilon_{xy}^0 = 1 \times 10^{-5}$, $0.5 \times 10^{-5}$ and $0$ for *k* changing from *X* to *Y'*. Note that the opposite sign of $\varepsilon_{xx}^0$ between the $k \parallel X$ and $k \parallel Y'$ geometries originates from the phase shift of longitudinal strains [57]. Figure 7 presents $V_{xx}^{\mathrm{AOHE}}$ as a function of the magnetic field angle $\varphi$. Clearly, $V_{xx}^{\mathrm{AOHE}}$ is enhanced by including a finite horizontal-shear strain $\varepsilon_{xy}$, which agrees with our experimental results in Section III.

Besides the acoustic OHE, the longitudinal voltage associated with the acoustic orbital pumping is also derived and discussed in Appendix C, taking the shear strains into consideration.

### C. Alternative mechanisms

#### 1. SAW-induced evanescent electromagnetic field

An alternative explanation of the acoustic OHE can be given in terms of the electromagnetic evanescent field associated with SAWs in the piezoelectric substrate, as recently proposed for the acoustic SHE [66,67]. The electric field component of the evanescent wave can drive an alternate charge current along *k* in the nonmagnetic layer, inducing an alternate spin or orbital current through either the SHE or OHE. In the present study, we cannot exclude such a contribution to the acoustic OHE. However, the efficiency of the acoustic OHE in Ti/Ni is larger compared to that of the acoustic SHE in Pt/Ni (Fig. 2), despite Ti having a significantly smaller orbital Hall conductivity [14,20]. This observation conflicts with the evanescent electric field model, which predicts that the acoustic voltage scales with the square of the spin or orbital Hall conductivity [66,67]. Thus, the evanescent electric field is unlikely to be the dominant cause of the acoustic OHE.

#### 2. Spin-rotation coupling

An ac spin current can be generated by the coupling between spins and mechanical rotation of the lattice, so-called spin–vorticity or spin-rotation coupling [68]. In a system where the lattice is subject to a mechanical rotation due to acoustic waves, the electrons feel a "vorticity field" proportional to the local lattice velocity. According to relativistic quantum mechanics, the spin couples to the angular velocity of the local rotation in the same way as it couples to a magnetic field. This mechanism of spin-current generation is independent of SOC, and the generated spin current is proportional to the spin lifetime. In this respect, light metals with long spin relaxation times are favorable systems for generating a spin current by spin rotation. However, the spin current from spin-rotation coupling scales with the third (fourth) power of the SAW's frequency $\omega$ for spin relaxation times greater (smaller) than $\omega^{-1}$ [68]. This is at odds with the linear dependence of the acoustic OHE voltage reported in Section 2. Moreover, the calculated spin current in Pt from spin-rotation coupling is of the order of $10^6$ A/m$^2$ at a frequency of 10 GHz. Therefore, such a contribution is expected to be negligible at a frequency of 460 MHz. We thus rule out the spin-rotation coupling effect in our experiments.

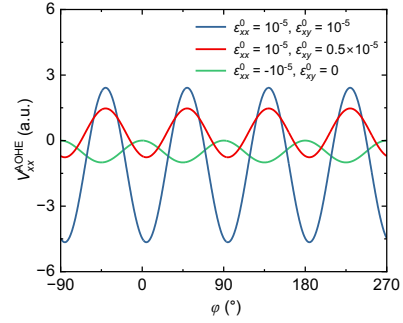

FIG. 7. Simulations of $V_{xx}^{\mathrm{AOHE}}$ as a function of the parameters of $\varepsilon_{xx}^0$, $\varepsilon_{xy}^0$ using Eq. (A19).

#### 3. Spin Hall effect caused by linear electron acceleration

Theoretical calculations show that the spin current can be generated from the linear acceleration of electrons in a vibrating lattice [69]. Under the SAWs, the longitudinal vibrations of the atoms induce an alternating electric field that generates an orthogonal spin current via the spin Hall effect. A similar mechanism can be envisaged for an orbital current in the presence of the OHE. However, the generated spin or orbital current should scale as $\omega^2$ and is also predicted to be small, about $10^5$ A/m$^2$ at a frequency of 10 GHz [69]. Therefore, this contribution can be ruled out in our measurements

#### 4. Thermal effects induced by SAWs

SAWs propagating in devices may induce a temperature increase in the film which is negligibly small. It may cause a longitudinal voltage through magnetothermal effects [45]. Specifically, in nonmagnetic/ferromagnetic bilayers, a longitudinal temperature gradient may generate a longitudinal voltage through the spin Nernst effect [70] and spin Seebeck effect [71]. However, these voltages exhibit in-plane magnetic field dependence proportional to $\sin^2\varphi$ and $\sin\varphi$, respectively. Additionally, a vertical temperature gradient may generate a longitudinal voltage through the anomalous Nernst effect, which would contribute a term proportional to $\sin\varphi$ in the in-plane magnetic field scans. The different $\varphi$–dependence of the measured acoustic voltage shows that thermal effects are negligible in all of our samples.

## VI. CONCLUSIONS

In summary, the acoustic OHE emerges prominently in light 3d metals such as Ti and Cr, driven by phonon–orbital coupling under SAW excitation. The effect is detected through rectification of the orbital current via magnetoelastic coupling in an adjacent ferromagnet, combined with its conversion into a charge signal by the inverse OHE in the light metal. The acoustic OHE is strongest at the SAW's resonant frequency, yet does not require resonant excitation of the magnetization. Conversely, the acoustic excitation of FMR in the ferromagnetic layer via magnetoelastic coupling results in orbital pumping into the nonmagnetic light metal. The rectified voltages from the acoustic OHE and orbital pumping can be distinguished by their different symmetry and magnetic field dependence.

Our work shows that the acoustic OHE occurs in diverse combinations of nonmagnetic and ferromagnetic materials. For a given nonmagnetic metal layer, the effect scales with the magnetoelastic coupling and orbital-spin conversion efficiency of the ferromagnetic layer. We also show that the acoustic OHE in Ti/Ni exceeds the acoustic SHE in Pt/Ni, which conflicts with the evanescent electric field model proposed in Refs. [66,67]. In addition, measurements of the acoustic OHE with varying SAW propagation directions relative to the LiNbO$_3$ substrate evidence a substantial contribution of horizontal-shear strain to the rectified acoustic voltages. Based on these results, we propose a phenomenological model of the acoustic OHE based on phonon–orbital coupling. The model accounts for the symmetry of both orbital Hall and pumping voltages, providing a framework for understanding orbital effects induced by lattice dynamics.

Our findings imply a highly effective phonon–orbital coupling in thin metal films, opening new opportunities to generate and modulate angular momentum currents through phonons. SAWs propagate over mm-scale distances without producing Joule heating. Thus, the acoustic OHE may be used to remotely manipulate and sense the magnetization by excitation of orbital degrees of freedom.

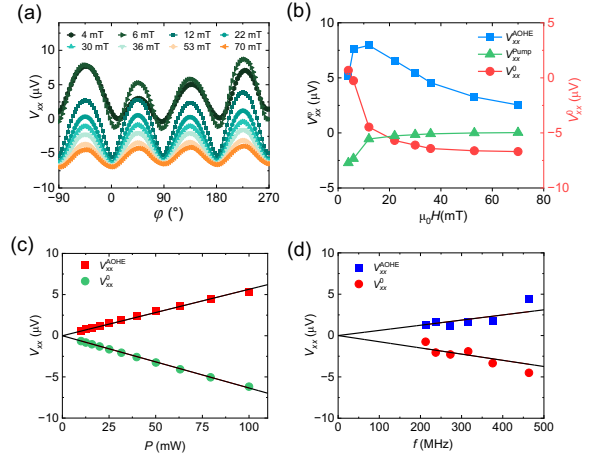

FIG. A1. Additional characterizations of acoustic OHE. (a) Angular profiles of $V_{xx}$ under different magnetic fields. (b) Amplitudes $V_{xx}^0$, $V_{xx}^{\text{AOHE}}$ and $V_{xx}^{\text{Pump}}$ as a function of magnetic field. Error bars from the standard deviation of the fits are smaller than the symbol size. (c) Dependence of $V_{xx}^{\text{AOHE}}$ and $V_{xx}^0$ on the input power fed into IDT1 for a Ti(20)/Ni(6) device with $k$ // $X$. (d) Frequency dependence of the acoustic OHE. Several Ti(20)/Ni(6) devices with $k \parallel X$ are designed with wavelengths from 8 to 18 μm. The magnetic field is set to 30 mT for the measurements in(c) and (d). Error bars are standard deviations from the fits.

## ACKNOWLEDGMENTS

This work was funded by the Swiss National Science Foundation (Grant No. 200021-236524) and JSPS KAKENHI (Grant No. 23KJ1159). H. M. acknowledges the support of the Swiss Government Excellence Scholarship for the 2024-2025 Academic Year.

The supporting data for this article are openly available from the ETH Research Collection https://doi.org/10.3929/ethz-b-000707406.

## APPENDIX A: Additional characterizations of the acoustic OHE

We further characterized the acoustic OHE by measuring $V_{xx}$ as a function of magnetic field. Figure A1(a) shows the results measured in a Ti(20)/Ni(6) device with $k \parallel X$. We found that $V_{xx}$ deviates from the four-fold symmetry at small magnetic fields. Such a deviation originates from the additional acoustic orbital pumping contribution, which becomes larger upon approaching the ferromagnetic resonance condition of the Ni layer. It is significantly suppressed by increasing the magnetic field. By fitting $V_{xx}$ with Eq. (3), $V_{xx}$ can be decomposed into three components: $V_{xx}^0$, $V_{xx}^{\text{AOHE}}$ and $V_{xx}^{\text{Pump}}$. Their magnitudes are plotted in Fig. A1(b) as a function of magnetic field $\mu_0 H$. The absolute magnitudes of $V_{xx}^0$ and $V_{xx}^{\text{Pump}}$ monotonically decrease with increasing magnetic field as the SAW-induced oscillations of the magnetization are gradually suppressed. On the other hand, $V_{xx}^{\text{AOHE}}$ increases at a small magnetic field and reaches a maximum at 12 mT, as the magnetic domains are fully aligned under this magnetic field. $V_{xx}^{\text{AOHE}}$ then decreases when $\mu_0 H > 12$ mT as the magnitude of the acoustic OHE is inversely proportional to the magnetic field. Therefore, we applied a magnetic field of 30 mT to saturate magnetization when studying the acoustic OHE and minimize contributions due to $V_{xx}^0$ and $V_{xx}^{\text{Pump}}$.

In the experiments, we applied an input power of 20 dBm (100 mW) to enhance the signal-to-noise ratio. Previous work revealed a nonlinear power dependence of the acoustic FMR excited by SAWs at 3.03 GHz [72]. This effect originates from the frequency shift of the SAW's resonance peaks due to heating near the IDTs. This nonlinearity is sensitive to the designed frequency of IDTs and becomes more pronounced for IDTs working at higher frequency [72,73]. To check for this effect in our devices, we measured the acoustic OHE as a function of the input power in Ti(20)/Ni(6), as shown in Fig. A1(c). We found that both $V_{xx}^{\text{AOHE}}$ and $V_{xx}^0$ scale proportionally to the input power, with no observable nonlinearity in the range of 10 to 100 mW. This suggests the negligible heating effect with the SAWs at ~ 460 MHz. It is consistent with previous work on SAWs [74], where the SAW-induced temperature increase was found to be smaller than 0.4 K. The variation of acoustic OHE efficiency, quantified by the ratio $-V_{xx}^{\text{AOHE}}/V_{xx}^0$, is ~ 7% from 10 to 100 mW. We therefore conclude that the heating-induced nonlinearities are small at 100 mW and can be safely neglected in our analysis.

We designed SAW devices with different spacings of the IDT fingers to investigate the frequency dependence of the acoustic OHE. Figure A1(d) presents $V_{xx}^{\text{AOHE}}$ and $V_{xx}^0$ in different Ti(20)/Ni(6) devices as a function of the SAW's resonance frequency. Both factors scale linearly with the resonance frequency. Therefore, we can rule out contributions to the acoustic voltage due to spin-rotation coupling, which are expected to be proportional to a higher order of the SAW's frequency [68]. We notice that SAW's transmission can vary from device to device due to the IDT fabrication, leading to variations of $V_{xx}^{\text{AOHE}}$ and $V_{xx}^0$. We therefore normalize $V_{xx}^{\text{AOHE}}$ by $V_{xx}^0$ to eliminate variations in SAW's transmission when comparing the acoustic OHE between different devices.

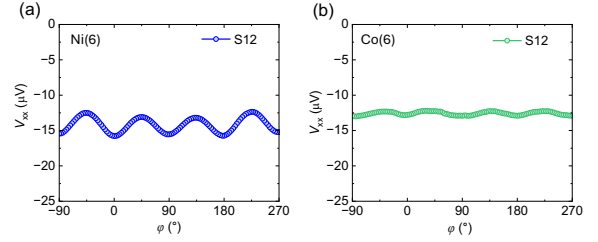

FIG. A2. Self-induced acoustic voltages in (a) Ni(6) and (b) Co(6) devices with $k \parallel X$. The applied magnetic field is 30 mT.

## APPENDIX B: Acoustic signals in single ferromagnetic layers

We found that the acoustic signals are nonvanishing even in single ferromagnetic layers. Figure A2 shows the angular dependence of acoustic voltages in single-layer Ni(6) and Co(6) devices under a magnetic field of 30 mT. Ni is well-known to exhibit self-generated torques due to either the OHE or spin Hall effect [20,32]. These two effects can cause finite self-induced acoustic Hall and pumping signals, as shown in Fig. A2(a). We obtain $-V_{xx}^{\text{AOHE}}/V_{xx}^0 = 0.18$ and $I_{xx} = 12$ nA for the Ni(6) single layer, which explains the nonzero offsets of $C_1$ and $C_2$ in Figs. 2(a) and 6. On the other hand, Co shows significantly smaller self-torques than Ni. Consistent with this, the self-induced acoustic signal is much smaller in Co, as shown in Fig. A2(b).

## APPENDIX C: Model of the acoustic OHE and acoustic orbital pumping

### 1. Phonon–orbital coupling

In this section, we establish a phenomenological model for the interaction between phonons and orbitals, extending the phenomenological theory of the acoustic SHE [45] to the orbital case and taking into account both longitudinal and shear strain effects. In the experiment, the phonons are carried by SAWs. The

SAW-induced strains $\varepsilon_{ij}$ can be expressed from the lattice displacement $u$ as:

$$\varepsilon_{ij} = \frac{\partial_j u_i + \partial_i u_j}{2}, \quad (A1)$$

where $i, j = x, y, z$. We defined the $x$-axis as always parallel to the SAW propagation direction. Eq. (A1) thus includes longitudinal strains when $i = j$ and shear strains when $i \neq j$. Due to the coupling of orbitals and crystal field, phononic excitations distort the orbital texture of a material, leading to so-called orbital torsion and oscillating orbital angular momenta [41,42]. To account for this effect, we define a dimensionless coefficient $\chi_{\text{po}}$ that defines phonon–orbital coupling. This coefficient plays an analogous role to the SOC in the acoustic SHE [45]. However, phonon–orbital coupling is independent of SOC and $\chi_{\text{po}}$ can be considerably stronger than SOC given the relative scale of crystal-field and spin-orbit interactions [43]. With this assumption, and including symmetry constraints analogous to Ref. [45], the orbital current density stems from the time derivative of the lattice displacements as:

$$j_{o,i}^{\zeta} = -\chi_{\text{po}} n e \epsilon_{\zeta ij} \frac{\partial u_j}{\partial t}, \quad (A2)$$

where $n$ is the electron density, $e$ the electron charge, and $\epsilon_{\zeta ij}$ the Levi-Civita symbol where $\zeta, i, j = x, y, z$. In this picture, the orbital polarization ($\zeta$) is orthogonal to the direction of the orbital current ($i$) and the lattice displacement ($j$). The lattice displacement $u_j$ can be expressed as a plane wave $u_j = u_j^0 e^{i(kx - \omega t)}$, where $u_j^0$ is the wave amplitude, $k$ the wavevector and $\omega$ the angular frequency. Inserting $u_j$ into Eq. (A2), we obtain an oscillating orbital current

$$j_{o,i}^{\zeta} = i\omega \chi_{\text{po}} n e \epsilon_{\zeta ij} u_j^0 e^{i(kx - \omega t)}. \quad (A3)$$

This expression shows that longitudinal lattice oscillations along the SAW's propagation direction $x$ lead to a transverse orbital current propagating along $z$ with polarization parallel to $y$, as expected of the OHE.

## 2. Acoustic orbital current rectification by magnetoelastic coupling

Direct detection of the acoustic orbital current given by Eq. (A3) is challenging due to its oscillating nature. However, such a current can be rectified by reflection from a ferromagnetic material with magnetization oscillating at the same frequency $\omega$. Here we modify the model introduced in the first study of the acoustic SHE [45] to include phonon–orbital coupling and shear-horizontal strain $\varepsilon_{xy}$ contributions. Rectification occurs upon reflection of the orbital current injected from a nonmagnetic material into a magnetic layer with magnetization aligned with the orbital polarization direction. The reflection process is the same as the one leading to spin Hall and orbital Hall magnetoresistances [49,50,75]. In SAWs-based experiments of nonmagnetic-ferromagnetic bilayers, the magnetization adiabatically oscillates at frequency $\omega$ due to magnetoelastic coupling [45,62], leading to a rectified orbital current being reflected back into the nonmagnetic layer. The back-reflected orbital current is then converted by the inverse OHE in charges accumulating at the edges of the sample, leading to a dc longitudinal voltage $V_{xx}$.

For SAWs propagating along the $x$-direction, only the longitudinal strain $\varepsilon_{xx}$ and shear strains $\varepsilon_{xy}$, $\varepsilon_{xz}$ are relevant for the magnetoelastic coupling effect. The magnetoelastic energy density of a polycrystalline film reads:

$$E_{\text{ME}} = B(\varepsilon_{xx} m_x^2 + 2\varepsilon_{xy} m_x m_y + 2\varepsilon_{xz} m_x m_z). \quad (A4)$$

Here, $\mathbf{m} = (m_x, m_y, m_z)$ is the unit vector of the magnetization and, for simplicity, we consider an isotropic material with magnetoelastic constant $B = B_1 = B_2$. The effective magnetic field $\mathbf{h}_{\text{eff}}$ due to magnetoelastic coupling is thus

$$\mathbf{h}_{\text{eff}} = -\frac{1}{\mu_0 M_s} \nabla E_{\text{ME}}, \quad (A5)$$

where $\mu_0$ is the vacuum permeability and $M_s$ the saturation magnetization. To obtain the analytical expression of $\mathbf{h}_{\text{eff}}$, we substitute the magnetization in spherical coordinates $\mathbf{m} = (\sin\theta\cos\varphi, \sin\theta\sin\varphi, \cos\theta)$ in Eq. (A4), and get:

$$h_{\text{eff}}^{\theta} = -\frac{1}{\mu_0 M_s} \frac{\partial E_{\text{ME}}}{\partial \theta} = -\frac{B}{\mu_0 M_s}\left(\varepsilon_{xx}\sin 2\theta \cos^2\varphi + \varepsilon_{xy}\sin 2\theta \sin 2\varphi + 2\varepsilon_{xz}\cos 2\theta \cos\varphi\right), \quad (A6)$$

and

$$h_{\text{eff}}^{\varphi} = -\frac{1}{\mu_0 M_s \sin\theta} \frac{\partial E_{\text{ME}}}{\partial \varphi} = -\frac{B}{\mu_0 M_s}\left(-\varepsilon_{xx}\sin\theta\sin 2\varphi + 2\varepsilon_{xy}\sin\theta\cos 2\varphi - 2\varepsilon_{xz}\cos\theta\sin\varphi\right), \quad (A7)$$

where $h_{\text{eff}}^\theta$ and $h_{\text{eff}}^\varphi$ are the projections of $\boldsymbol{h}_{\text{eff}}$ along the polar and azimuthal unit vectors, respectively. Owing to the oscillating nature of $\boldsymbol{h}_{\text{eff}}$ under SAWs' excitation, the magnetization oscillates around the equilibrium position defined by the external magnetic field and magnetic anisotropy. Writing $\boldsymbol{m}$ as $(m_\theta, m_\varphi, m_z)$ and considering the small oscillations limit $m_\theta, m_\varphi \ll 1$ and $m_z \approx 1$, the components $(m_\theta, m_\varphi)$ can be deduced from the linearized Landau-Lifshitz-Gilbert equation after Fourier transforms in time and space [76]:

$$\begin{pmatrix} m_\theta \\ m_\varphi \end{pmatrix} = \frac{1}{H^2 - H_\omega^2 - (\alpha H_\omega)^2 + 2i\alpha H_\omega H} \times \begin{pmatrix} H + i\alpha H_\omega & -iH_\omega \\ iH_\omega & H + i\alpha H_\omega \end{pmatrix} \begin{pmatrix} h_{\text{eff}}^\theta \\ h_{\text{eff}}^\varphi \end{pmatrix}. \text{(A8)}$$

Here, $H_\omega = \omega/\gamma\mu_0$ where $\gamma$ is the gyromagnetic ratio, and $\alpha$ is the Gilbert damping constant. To simplify the equation, we discard the $\alpha$ terms due to the small magnetic damping of Ni ( $\alpha \sim 0.04$ in Ti/Ni bilayer [32]). Then Eq. (A8) becomes:

$$\begin{pmatrix} m_\theta \\ m_\varphi \end{pmatrix} = \frac{1}{H^2 - H_\omega^2} \begin{pmatrix} H & -iH_\omega \\ iH_\omega & H \end{pmatrix} \begin{pmatrix} h_{\text{eff}}^\theta \\ h_{\text{eff}}^\varphi \end{pmatrix}. \quad \text{(A9)}$$

To derive the rectified acoustic OHE voltage, we use a drift-diffusion model analogous to that employed to describe the spin Hall magnetoresistance and acoustic SHE in nonmagnetic/ferromagnetic bilayers [49]. Orbital diffusion in the nonmagnetic layer is thus described by the diffusion equation:

$$\nabla^2 \boldsymbol{\mu} = \frac{\boldsymbol{\mu}}{\lambda_N^2}, \quad \text{(A10)}$$

where $\boldsymbol{\mu}$ is the orbital chemical potential and $\lambda_N$ the orbital diffusion length. For a bilayer extending in the $x$-$y$ plane, the acoustic voltage arises from the orbital current flowing along $z$. Therefore, only the orbital polarization in the $x$-$y$ plane ($\zeta = x, y$) contributes to the observed acoustic OHE voltage. Thus, the orbital polarization can be written as $\boldsymbol{l} = \eta\boldsymbol{x} + \boldsymbol{y}$, where $\eta$ is the ratio of the orbital polarization along $x$ and $y$. Thus, the lattice displacements $u_x$, $u_y$ are responsible for the orbital polarization along $y$ and $x$, respectively. The orbital current density flowing along $z$ reads:

$$\boldsymbol{j}_{\text{o},z} = -\frac{\sigma}{2e}\partial_z\boldsymbol{\mu} - j_{\text{o},z}^l \boldsymbol{l}, \quad \text{(A11)}$$

where $j_{\text{o},z}^l = \omega\chi_{\text{po}}ne[(u_x^0)^2 + (u_y^0)^2]^{\frac{1}{2}}$ is the orbital current density generated directly by the acoustic OHE and $\sigma$ the electrical conductivity. By solving Eq. (A11) along $z$, we find $\boldsymbol{\mu}(z) = \boldsymbol{A}e^{\frac{z}{\lambda_N}} + \boldsymbol{B}e^{-\frac{z}{\lambda_N}}$ where $\boldsymbol{A}$, $\boldsymbol{B}$ are parameters determined by the boundary conditions at the nonmagnetic layer's interface, and $z = 0$ defines the interface between nonmagnetic material and piezoelectric substrate. The orbital current generated in the nonmagnet results in orbital accumulation at the interface with the magnetic layer ($z = t_N$), whereby the orbital current is either reflected or absorbed by the ferromagnet depending on the relative orientation of $\boldsymbol{l}$ and $\boldsymbol{m}$. Under these conditions, $\boldsymbol{j}_{\text{o},z}$ vanishes at $z = 0$, and is continuous at the NM/FM interface. Thus, the orbital current in the nonmagnetic layer has the following profile [45,49]:

$$\boldsymbol{j}_{\text{o},z} = j_{\text{o},z}^l \{c_1\boldsymbol{l} - c_2[\boldsymbol{m}\times(\boldsymbol{m}\times\boldsymbol{l})\text{Re} - (\boldsymbol{m}\times\boldsymbol{l})\text{Im}]\frac{2\lambda_N G_\pm}{\sigma + 2\lambda_N G_\pm}\}, \quad \text{(A12)}$$

where $c_1 = \frac{\cosh\left(\frac{2z-t_N}{2\lambda_N}\right) - \cosh\left(\frac{t_N}{2\lambda_N}\right)}{\cosh\left(\frac{t_N}{2\lambda_N}\right)}$ and $c_2 = -\frac{\tanh\left(\frac{t_N}{2\lambda_N}\right)}{\coth\left(\frac{t_N}{\lambda_N}\right)} \frac{\sin\left(\frac{z}{\lambda_N}\right)}{\sin\left(\frac{t_N}{\lambda_N}\right)}$ are the diffusion profiles, and $G_\pm$ is the interfacial orbital-mixing conductance. We note that $G_\pm$ represents an effective parameter akin to the complex spin-mixing conductance, whose real and imaginary parts model the absorption and rotation of $\boldsymbol{l}$ by the ferromagnet, respectively. A detailed treatment of these effects for an orbital current should include different mixing conductances for different orbitals. Here, we consider an ideal interface limit, where the transverse spin current is fully absorbed at the NM/FM boundary, such that the real part of the spin-mixing conductance dominates the spin conductance and $\text{Re}[G_\pm] \gg \frac{\sigma}{2\lambda_N}$.

To obtain an explicit expression for the orbital current, we find $\boldsymbol{m}$ by solving Eq. (A9). In the measurements of the acoustic OHE we apply a magnetic field $H = 30$ mT $\gg H_\omega$, where $H_\omega \approx 1.5$

mT at the frequency of the SAW ($\approx 460$ MHz). Therefore, Eq. (A9) simplifies as:

$$\begin{pmatrix} m_\theta \\ m_\varphi \end{pmatrix} = \begin{pmatrix} 1/H & 0 \\ 0 & 1/H \end{pmatrix} \begin{pmatrix} h_{\text{eff}}^\theta \\ h_{\text{eff}}^\varphi \end{pmatrix}. \quad (A13)$$

For a magnetic field in the $x$-$y$ plane (i.e, $\theta = \pi/2$) and inserting Eqs. (A6) and (A7) into Eq. (A13), we obtain:

$$\begin{pmatrix} m_\theta \\ m_\varphi \end{pmatrix} = -\frac{B}{\mu_0 HM_s} \begin{pmatrix} -2\varepsilon_{xz}\cos\varphi \\ -\varepsilon_{xx}\sin 2\varphi + 2\varepsilon_{xy}\cos 2\varphi \end{pmatrix}. \quad (A14)$$

By converting $\boldsymbol{m}$, including the dynamic components, back in Cartesian coordinates, we have:

$$\begin{pmatrix} m_x \\ m_y \\ m_z \end{pmatrix} = \begin{pmatrix} \cos\varphi - \frac{B}{\mu_0 HM_s}(\varepsilon_{xx}\sin 2\varphi \sin\varphi - 2\varepsilon_{xy}\cos 2\varphi \sin\varphi) \\ \sin\varphi - \frac{B}{\mu_0 HM_s}(-\varepsilon_{xx}\sin 2\varphi \cos\varphi + 2\varepsilon_{xy}\cos 2\varphi \cos\varphi) \\ -\frac{2b}{\mu_0 HM_s}\varepsilon_{xz}\cos\varphi \end{pmatrix}. \quad (A15)$$

Inserting Eq. (A15) into Eq. (A12) and averaging the orbital current in time and space, we get the DC component of the orbital current:

$$\langle \boldsymbol{j}_{o,z} \rangle = j_{o,z}^l \langle c_2 \rangle \frac{B}{\mu_0 HM_s} \begin{pmatrix} \varepsilon_{xx}^0 (\eta \sin^2 2\varphi - \frac{\sin 4\varphi}{2}) + \varepsilon_{xy}^0 (2\cos^2 2\varphi - \eta \sin 4\varphi) \\ \varepsilon_{xx}^0 \left(\sin^2 2\varphi + \frac{\eta \sin 4\varphi}{2}\right) - \varepsilon_{xy}^0 (2\eta \cos^2 2\varphi + \sin 4\varphi) \\ 0 \end{pmatrix}, \quad (A16)$$

where $\langle c_2 \rangle = \frac{1}{t_N}\int_0^{t_N} c_2 dz = -\frac{\lambda_N}{t_N}\tanh^2\left(\frac{t_N}{2\lambda_N}\right)\tanh\left(\frac{t_N}{\lambda_N}\right)$ represents the averaged $c_2$ along $z$, describing the orbital diffusion, and accounts for Eq. (1) in the main text. The factors $\varepsilon_{xx}^0 = |k|u_x^0$ and $\varepsilon_{xy}^0 = \frac{|k|}{2}u_y^0$ are the amplitudes of longitudinal and shear-horizontal strains.

The orbital current is converted into a charge current via the inverse OHE. The charge current is expressed as:

$$\langle \boldsymbol{j}_c \rangle = \theta_{\text{OHE}}\langle \boldsymbol{j}_{o,z} \rangle \times \boldsymbol{z} = \theta_{\text{OHE}}\langle c_2 \rangle j_{o,z}^l \frac{B}{\mu_0 HM_s} \begin{pmatrix} \varepsilon_{xx}^0 \left(\sin^2 2\varphi + \frac{\eta \sin 4\varphi}{2}\right) - \varepsilon_{xy}^0 (2\eta \cos^2 2\varphi + \sin 4\varphi) \\ -\varepsilon_{xx}^0 \left(\eta \sin^2 2\varphi - \frac{\sin 4\varphi}{2}\right) - \varepsilon_{xy}^0 (2\cos^2 2\varphi - \eta \sin 4\varphi) \\ 0 \end{pmatrix}. \quad (A17)$$

Inserting $j_{o,z}^l$ into Eq. (A17), we obtain the $x$-component of charge current as:

$$\langle j_{c,x} \rangle = \theta_{\text{OHE}}\chi_{\text{po}}ne\frac{\omega}{|k|}\langle c_2 \rangle \frac{B}{\mu_0 HM_s}[(\varepsilon_{xx}^0)^2 + (2\varepsilon_{xy}^0)^2]^{\frac{1}{2}}\left[\varepsilon_{xx}^0\left(\sin^2 2\varphi + \frac{\eta \sin 4\varphi}{2}\right) - \varepsilon_{xy}^0(2\eta \cos^2 2\varphi + \sin 4\varphi)\right]. \quad (A18)$$

For a device with resistivity $\rho_N$ and length $L$, the longitudinal voltage induced by acoustic OHE reads:

$$V_{xx} = \rho_N L \theta_{\text{OHE}}\chi_{\text{po}}ne\frac{\omega}{|k|}\langle c_2 \rangle \frac{B}{\mu_0 HM_s}[(\varepsilon_{xx}^0)^2 + (2\varepsilon_{xy}^0)^2]^{\frac{1}{2}}\left[\varepsilon_{xx}^0\left(\sin^2 2\varphi + \frac{\eta \sin 4\varphi}{2}\right) - \varepsilon_{xy}^0(2\eta \cos^2 2\varphi + \sin 4\varphi)\right]. \quad (A19)$$

Eq. (A19) has a format of $V_{xx} = V_{xx}^{\text{AOHE}}\sin^2(2\varphi + \delta) + V_{xx}^{\text{offset}}$, with $V_{xx}^{\text{AOHE}} = \Gamma(1+\eta^2)^{\frac{1}{2}}[(\varepsilon_{xx}^0)^2 + (2\varepsilon_{xy}^0)^2]$ where $\Gamma = \rho_N L \theta_{\text{OHE}}\chi_{\text{po}}ne\frac{\omega}{|k|}\langle c_2 \rangle \frac{B}{\mu_0 HM_s}$ and $\tan 2\delta = -\frac{\eta \varepsilon_{xx}^0 - 2\varepsilon_{xy}^0}{\varepsilon_{xx}^0 + 2\eta \varepsilon_{xy}^0}$. Thus, our model can well interpret Eq. (2) for the acoustic OHE voltage in the main text.

### 3. Longitudinal voltage from acoustic orbital pumping

Acoustic orbital pumping, analogously to acoustic spin pumping [62], becomes prominent when FMR is excited in the ferromagnetic layer, resulting in an orbital momentum current being pumped out of the ferromagnet and converted into a charge current in the nonmagnetic layer. To derive the orbital pumping due to acoustic FMR, we need to consider $H_\omega$ in Eq. (A9), which gives:

$$\begin{pmatrix} m_\theta \\ m_\varphi \end{pmatrix} = \frac{1}{H^2 - H_\omega^2} \begin{pmatrix} H h_{\text{eff}}^\theta - i H_\omega h_{\text{eff}}^\varphi \\ i H_\omega h_{\text{eff}}^\theta + H h_{\text{eff}}^\varphi \end{pmatrix}. \quad (A20)$$

The orbital current density from the orbital pumping is expressed as [61]:

$$\boldsymbol{j}_{o,z} = \frac{\hbar}{4\pi} \text{Re}\left[\frac{2\lambda_N G_\pm}{\sigma + 2\lambda_N G_\pm}\right] (\boldsymbol{m} \times \partial_t \boldsymbol{m}). \quad (A21)$$

The DC component of $j_{o,z}$ is obtained by time averaging of Eq. (A21), giving

$$\langle j_{o,z} \rangle = \frac{\hbar \omega}{8\pi} \text{Re}\left[\frac{2\lambda_N G_\pm}{\sigma + 2\lambda_N G_\pm}\right] \text{Im}[m_\theta^* m_\varphi - m_\theta m_\varphi^*], \quad (A22)$$

where the orbital polarization is parallel to the static component of the magnetization. Substituting $m_\theta$ and $m_\varphi$ in Eq. (A22) and taking $\text{Re}[\frac{2\lambda_N G_\pm}{\sigma + 2\lambda_N G_\pm}] \approx 1$ by assuming $G_\pm \gg \sigma/2\lambda_N$, gives:

$$\langle j_{o,z} \rangle = \frac{\hbar \omega H H_\omega B^2}{(H^2 - H_\omega^2)^2 \mu_0^2 M_s^2} \left[(\varepsilon_{xx}^0)^2 \sin^2 2\varphi + (\varepsilon_{xy}^0)^2 \cos^2 2\varphi + 4(\varepsilon_{xz}^0)^2 \cos^2 \varphi\right]. \quad (A23)$$

This pumped orbital current is then converted into a charge current through the inverse OHE of the nonmagnetic layer. Taking into account orbital diffusion within the nonmagnetic layer with the same boundary conditions as introduced above, and averaging the charge current in space, we get the longitudinal component of the charge current density:

$$\langle j_{c,x} \rangle = \frac{c_3 \theta_{\text{OHE}} \hbar \omega H H_\omega B^2}{(H^2 - H_\omega^2)^2 \mu_0^2 M_s^2} \left[(\varepsilon_{xx}^0)^2 \sin^2 2\varphi + (\varepsilon_{xy}^0)^2 \cos^2 2\varphi + 4(\varepsilon_{xz}^0)^2 \cos^2 \varphi\right] \sin\varphi, \quad (A24)$$

where $c_3 = \frac{t_N}{\lambda_N} \tanh \frac{t_N}{2\lambda_N}$ is the diffusion profile of the pumped orbital current. The shear strain $\varepsilon_{xz}$ is usually smaller than the longitudinal strain $\varepsilon_{xx}$ and is only sizeable in thick films [77]. We thus ignore $\varepsilon_{xz}$ and obtain the orbital pumping voltage contribution in Eq. (3) of the main text as:

$$V_{xx} = \rho_N L \frac{c_3 \theta_{\text{OHE}} \hbar \omega H H_\omega B^2}{(H^2 - H_\omega^2)^2 \mu_0^2 M_s^2} \left[(\varepsilon_{xx}^0)^2 \sin^2 2\varphi + (\varepsilon_{xy}^0)^2 \cos^2 2\varphi\right] \sin\varphi. \quad (A25)$$

The associated longitudinal voltage for the acoustic orbital pumping is given by $V_{xx}^{\text{pump}} = \rho_N L \frac{c_3 \theta_{\text{OHE}} \hbar \omega H H_\omega B^2}{(H^2 - H_\omega^2)^2 \mu_0^2 M_s^2} [(\varepsilon_{xx}^0)^2 - (\varepsilon_{xy}^0)^2]$. The amplitude of the pumping current in the nonmagnetic-ferromagnetic bilayer thus reads:

$$I_{xx}^{\text{Pump}} = \frac{|V_{xx}^{\text{Pump}}|}{R_{xx}} = A \frac{t_N}{\lambda_N} \tanh \frac{t_N}{2\lambda_N}, \quad (A26)$$

where $R_{xx}$ is the bilayer's resistance and $A = \frac{\rho_N L \theta_{\text{OHE}} \hbar \omega H H_\omega B^2 [(\varepsilon_{xx}^0)^2 - (\varepsilon_{xy}^0)^2]}{2(H^2 - H_\omega^2)^2 \mu_0^2 M_s^2 R_{xx}}$. Clearly, the acoustic orbital pumping signal is only pronounced when $H$ approaches the FMR field $H_\omega$.

[1] J. Sinova, S. O. Valenzuela, J. Wunderlich, C. H. Back, and T. Jungwirth, Spin Hall effects. Rev Mod Phys **87**, 1213–1260 (2015).
[2] B. A. Bernevig, T. L. Hughes, and S. C. Zhang, Orbitronics: The intrinsic orbital current in p-doped silicon, Phys Rev Lett **95**, 066601 (2005).
[3] H. Kontani, T. Tanaka, D. S. Hirashima, K. Yamada, and J. Inoue, Giant orbital hall effect in transition metals: Origin of large spin and anomalous hall effects, Phys Rev Lett **102**, 016601 (2009).
[4] D. Go, D. Jo, C. Kim, and H. W. Lee, Intrinsic spin and orbital Hall effects from orbital texture, Phys Rev Lett **121**, 086602 (2018).
[5] D. Jo, D. Go, and H. W. Lee, Gigantic intrinsic orbital Hall effects in weakly spin-orbit coupled metals, Phys Rev B **98**, 214405 (2018).
[6] L. Liu, C.-F. Pai, Y. Li, H. W. Tseng, D. C. Ralph, and R. A. Buhrman, Spin-torque switching with the giant spin Hall effect of tantalum, Science **336**, 555-558 (2012).
[7] I. M. Miron, K. Garello, G. Gaudin, P. J. Zermatten, M. V. Costache, S. Auffret, S. Bandiera, B. Rodmacq, A. Schuhl, and P. Gambardella, Perpendicular switching of a single ferromagnetic layer induced by in-plane current injection. Nature **476** 189–193 (2011).
[8] S. Lee et al., Efficient conversion of orbital Hall current to spin current for spin-orbit torque switching, Commun Phys **4**, 234 (2021).
[9] Z. C. Zheng et al., Magnetization switching driven by current-induced torque from weakly spin-orbit coupled Zr, Phys Rev Res **2**, 013127 (2020).
[10] Y. Yang et al., Orbital torque switching in perpendicularly magnetized materials, Nat Commun **15**, 8645 (2024).
[11] L. Salemi and P. M. Oppeneer, First-principles theory of intrinsic spin and orbital Hall and Nernst effects in metallic monoatomic crystals, Phys Rev Mater **6**, 095001 (2022).
[12] G. Sala and P. Gambardella, Giant orbital Hall effect and orbital-to-spin conversion in 3d, 5d, and 4f metallic heterostructures, Phys Rev Res **4**, 033037 (2022).
[13] D. Go, H. W. Lee, P. M. Oppeneer, S. Blügel, and Y. Mokrousov, First-principles calculation of orbital Hall effect by Wannier interpolation: Role of orbital dependence of the anomalous position, Phys Rev B **109**, 174435 (2024).
[14] Y. G. Choi, D. Jo, K. H. Ko, D. Go, K. H. Kim, H. G. Park, C. Kim, B. C. Min, G. M. Choi, and H. W. Lee, Observation of the orbital Hall effect in a light metal Ti, Nature **619**, 52 (2023).
[15] Y. Marui, M. Kawaguchi, S. Sumi, H. Awano, K. Nakamura, and M. Hayashi, Spin and orbital Hall currents detected via current-induced magneto-optical Kerr effect in V and Pt, Phys Rev B **108**, 144436 (2023).
[16] I. Lyalin, S. Alikhah, M. Berritta, P. M. Oppeneer, and R. K. Kawakami, Magneto-optical detection of the orbital Hall effect in chromium, Phys Rev Lett **131**, 156702 (2023).
[17] G. Sala, H. Wang, W. Legrand, and P. Gambardella, Orbital Hanle magnetoresistance in a 3d transition metal, Phys Rev Lett **131**, 156703 (2023).
[18] D. Lee et al., Orbital torque in magnetic bilayers, Nat Commun **12**, 6710 (2021).
[19] S. Ding, M. G. Kang, W. Legrand, and P. Gambardella, Orbital torque in rare-earth transition-metal ferrimagnets, Phys Rev Lett **132**, 236702 (2024).
[20] H. Hayashi, D. Jo, D. Go, Y. Mokrousov, H.-W. Lee, and K. Ando, Observation of long-range orbital transport and giant orbital torque, Commun Phys **6**, 32 (2022).
[21] M. Taniguchi, H. Hayashi, N. Soya, and K. Ando, Nonlocal orbital torques in magnetic multilayers, Appl Phys Express **16**, 043001 (2023).
[22] T. Gao et al., Control of dynamic orbital response in ferromagnets via crystal symmetry, Nat Phys **20**, 1896–1903 (2024).
[23] S. A. Nikolaev, M. Chshiev, F. Ibrahim, S. Krishnia, N. Sebe, J. M. George, V. Cros, H. Jaffres, and A. Fert, Large chiral orbital texture and orbital Edelstein effect in Co/Al heterostructure, Nano Lett **24**, 13465 (2024).
[24] S. Ding, H. Wang, W. Legrand, P. Noël, and P. Gambardella, Mitigation of Gilbert damping in the CoFe/CuO$_x$ orbital torque system, Nano Lett **24**, 10251 (2024).
[25] R. Gupta et al., Harnessing orbital Hall effect in spin-orbit torque MRAM, Nat Commun **16**, 130 (2025).
[26] S. Ding et al., Harnessing orbital-to-spin conversion of interfacial orbital currents for efficient spin-orbit torques, Phys Rev Lett **125**, 177201 (2020).
[27] T. Li, L. Liu, X. Li, X. Zhao, H. An, and K. Ando, Giant orbital-to-spin conversion for efficient current-induced magnetization switching of ferrimagnetic insulator, Nano Lett **23**, 7174 (2023).
[28] T. S. Seifert, D. Go, H. Hayashi, R. Rouzegar, F. Freimuth, K. Ando, Y. Mokrousov, and T. Kampfrath, Time-domain observation of ballistic orbital-angular-momentum currents with giant

[28] relaxation length in tungsten, Nat Nanotechnol **18**, 1132 (2023).
[29] Y. Xu, F. Zhang, A. Fert, H. Y. Jaffres, Y. Liu, R. Xu, Y. Jiang, H. Cheng, and W. Zhao, Orbitronics: light-induced orbital currents in Ni studied by terahertz emission experiments, Nat Commun **15**, 2043 (2024).
[30] P. Wang et al., Inverse orbital Hall effect and orbitronic terahertz emission observed in the materials with weak spin-orbit coupling, NPJ Quantum Mater **8**, 28 (2023).
[31] A. El Hamdi, J. Y. Chauleau, M. Boselli, C. Thibault, C. Gorini, A. Smogunov, C. Barreteau, S. Gariglio, J. M. Triscone, and M. Viret, Observation of the orbital inverse Rashba–Edelstein effect, Nat Phys **19**, 1855 (2023).
[32] H. Hayashi, D. Go, S. Haku, Y. Mokrousov, and K. Ando, Observation of orbital pumping, Nat Electron **7**, 646–652 (2024).
[33] H. Wang, M. G. Kang, D. Petrosyan, S. Ding, R. Schlitz, L. J. Riddiford, W. Legrand, and P. Gambardella, Orbital pumping in ferrimagnetic insulators, Phys Rev Lett **134**, 126701 (2025).
[34] N. Keller et al., Identification of orbital pumping from spin pumping and rectification effects, Nano Lett **25**, 13462 (2025).
[35] Y. Tserkovnyak, A. Brataas, G. E. W. Bauer, and B. I. Halperin, Magnetization dynamics in ferromagnetic heterostructures. Rev Mod Phys **77**, 1375 (2005).
[36] M. Basini, M. Pancaldi, B. Wehinger, M. Udina, V. Unikandanunni, T. Tadano, M. C. Hoffmann, A. V. Balatsky, and S. Bonetti, Terahertz electric-field-driven dynamical multiferroicity in $SrTiO_3$, Nature **628**, 534 (2024).
[37] C. S. Davies, F. G. N. Fennema, A. Tsukamoto, I. Razdolski, A. V. Kimel, and A. Kirilyuk, Phononic switching of magnetization by the ultrafast Barnett effect, Nature **628**, 540 (2024).
[38] D. T. Maimone, A. B. Christian, J. J. Neumeier, and E. Granado, Coupling of phonons with orbital dynamics and magnetism in $CuSb_2O_6$, Phys Rev B **97**, 174415 (2018).
[39] A. P. Roy et al., Evidence of strong orbital-selective spin-orbital-phonon coupling in $CrVO_4$, Phys Rev Lett **132**, 026701 (2024).
[40] C. Ulrich et al., Momentum dependence of orbital excitations in mott-insulating titanates, Phys Rev Lett **103**, 107205 (2009).
[41] S. Han, H. W. Lee, and K. W. Kim, Orbital dynamics in centrosymmetric systems, Phys Rev Lett **128**, 176601 (2022).
[42] S. Han, H. W. Ko, J. H. Oh, H. W. Lee, K. J. Lee, and K. W. Kim, Orbital pumping incorporating both orbital angular momentum and position, Phys Rev Lett **134**, 036305 (2025).
[43] K.-J. Lee, V. Cros, and H.-W. Lee, Electric-field-induced orbital angular momentum in metals, Nat Mater **23**, 1302 (2024).
[44] M. Taniguchi, S. Haku, H.-W. Lee, and K. Ando, Acoustic generation of orbital currents, Nat Commun **16**, 8038 (2025).
[45] T. Kawada, M. Kawaguchi, T. Funato, H. Kohno, and M. Hayashi, Acoustic spin Hall effect in strong spin-orbit metals, Sci Adv **7**, eabd9697 (2021).
[46] M. Rotter, A. Wixforth, W. Ruile, D. Bernklau, and H. Riechert, Giant acoustoelectric effect in $GaAs/LiNbO_3$ hybrids, Appl Phys Lett **73**, 2128 (1998).
[47] H. Moriya, M. Taniguchi, D. Jo, D. Go, N. Soya, H. Hayashi, Y. Mokrousov, H. W. Lee, and K. Ando, Observation of long-range current-induced torque in Ni/Pt bilayers, Nano Lett **24**, 6459 (2024).
[48] J. Kim, P. Sheng, S. Takahashi, S. Mitani, and M. Hayashi, Spin Hall magnetoresistance in metallic bilayers, Phys Rev Lett **116**, 097201 (2016).
[49] Y. T. Chen, S. Takahashi, H. Nakayama, M. Althammer, S. T. B. Goennenwein, E. Saitoh, and G. E. W. Bauer, Theory of spin Hall magnetoresistance, Phys Rev B **87**, 144411 (2013).
[50] H. Hayashi and K. Ando, Orbital Hall magnetoresistance in Ni/Ti bilayers, Appl Phys Lett **123**, 172401 (2023).
[51] Y. Hwang et al., Strongly coupled spin waves and surface acoustic waves at room temperature, Phys Rev Lett **132**, 056704 (2024).
[52] D. Go and H. W. Lee, Orbital torque: Torque generation by orbital current injection, Phys Rev Res **2**, 013177 (2020).
[53] D. Go, F. Freimuth, J. P. Hanke, F. Xue, O. Gomonay, K. J. Lee, S. Blügel, P. M. Haney, H. W. Lee, and Y. Mokrousov, Theory of current-induced angular momentum transfer dynamics in spin-orbit coupled systems, Phys Rev Res **2**, 033401 (2020).
[54] D. Hunter et al., Giant magnetostriction in annealed $Co_{1-x}Fe_x$ thin-films, Nat Commun **2**, 518 (2011).
[55] R. C. Hall, Single crystal anisotropy and magnetostriction constants of several ferromagnetic materials including alloys of NiFe, SiFe, AlFe, CoNi, and CoFe, J Appl Phys **30**, 816 (1959).
[56] E. Klokholm and J. Aboaf, The saturation magnetostriction of thin polycrystalline films of iron, cobalt, and nickel, J Appl Phys **53**, 2661 (1982).
[57] L. Liao, F. Chen, J. Puebla, J.-I. Kishine, K. Kondou, W. Luo, D. Zhao, Y. Zhang, Y. Ba, and

Y. Otani, Nonreciprocal magnetoacoustic waves with out-of-plane phononic angular momenta, Sci Adv **10**, eado2504 (2024).

[58] G. Tang, T. Han, J. Chen, B. Zhang, T. Omori, and K. Y. Hashimoto, Model parameter extraction for obliquely propagating surface acoustic waves in infinitely long grating structures. Jpn J Appl Phys **55**, 07KD08 (2016).

[59] B. Zhang, T. Han, G. Tang, Q. Zhang, T. Omori, and K. Y. Hashimoto, Influence of coupling with shear horizontal surface acoustic wave on lateral propagation of Rayleigh surface acoustic wave on 128°$YX$-LiNbO$_3$, Jpn J Appl Phys **56**, 07JD02 (2017).

[60] M. Weiler, L. Dreher, C. Heeg, H. Huebl, R. Gross, M. S. Brandt, and S. T. B. Goennenwein, Elastically driven ferromagnetic resonance in nickel thin films, Phys Rev Lett **106**, 117601 (2011).

[61] L. Dreher, M. Weiler, M. Pernpeintner, H. Huebl, R. Gross, M. S. Brandt, and S. T. B. Goennenwein, Surface acoustic wave driven ferromagnetic resonance in nickel thin films: Theory and experiment, Phys Rev B **86**, 134415 (2012).

[62] M. Weiler, H. Huebl, F. S. Goerg, F. D. Czeschka, R. Gross, and S. T. B. Goennenwein, Spin pumping with coherent elastic waves, Phys Rev Lett **108**, 176601 (2012).

[63] M. Xu, J. Puebla, F. Auvray, B. Rana, K. Kondou, and Y. Otani, Inverse Edelstein effect induced by magnon-phonon coupling, Phys Rev B **97**, 180301(R) (2018).

[64] K. Ando et al., Inverse spin-Hall effect induced by spin pumping in metallic system J Appl Phys **109**, 103913 (2011).

[65] T. Tanaka, H. Kontani, M. Naito, T. Naito, D. S. Hirashima, K. Yamada, and J. Inoue, Intrinsic spin Hall effect and orbital Hall effect in 4d and 5d transition metals, Phys Rev B **77**, 165117 (2008).

[66] T. Kawada, K. Yamamoto, M. Kawaguchi, H. Matsumoto, R. Hisatomi, H. Kohno, S. Maekawa, and M. Hayashi, Electromagnetic evanescent field associated with surface acoustic wave: Response of metallic thin films, arXiv:2412.13436. (2024).

[67] T. Kawada, M. Kawaguchi, K. Yamamoto, H. Matsumoto, R. Hisatomi, H. Kohno, S. Maekawa, and M. Hayashi, Spin current generation by acousto-electric evanescent wave, arXiv:2412.17291. (2024).

[68] M. Matsuo, J. Ieda, K. Harii, E. Saitoh, and S. Maekawa, Mechanical generation of spin current by spin-rotation coupling, Phys Rev B **87**, 180402(R) (2013).

[69] M. Matsuo, E. Saitoh, and S. Maekawa, Spin-mechatronics, Jpn J Appl Phys **86**, 011011 (2017).

[70] S. Meyer et al., Observation of the spin Nernst effect, Nat Mater **16**, 977 (2017).

[71] K. Uchida, S. Takahashi, K. Harii, J. Ieda, W. Koshibae, K. Ando, S. Maekawa, and E. Saitoh, Observation of the spin Seebeck effect, Nature **455**, 778 (2008).

[72] Y. Hwang, J. Puebla, K. Kondou, C. S. Muñoz, and Y. Otani, Nonlinear acoustic spin pumping caused by temperature-dependent frequency shifts of surface acoustic waves, Jpn J Appl Phys **92**, 094702 (2023).

[73] Y. Hwang, J. Puebla, M. Xu, A. Lagarrigue, K. Kondou, and Y. Otani, Enhancement of acoustic spin pumping by acoustic distributed Bragg reflector cavity, Appl Phys Lett **116**, 252404 (2020).

[74] T. Yokouchi, S. Sugimoto, B. Rana, S. Seki, N. Ogawa, S. Kasai, and Y. Otani, Creation of magnetic skyrmions by surface acoustic waves, Nat Nanotechnol **15**, 361 (2020).

[75] S. Ding et al., Observation of the orbital Rashba-Edelstein magnetoresistance, Phys Rev Lett **128**, 067201 (2022).

[76] M. Xu, K. Yamamoto, J. Puebla, K. Baumgaertl, B. Rana, K. Miura, H. Takahashi, D. Grundler, S. Maekawa, and Y. Otani, Nonreciprocal surface acoustic wave propagation via magneto-rotation coupling, Sci Adv **6**, eabb1724 (2020).

[77] S. Tateno, Y. Nozaki, and Y. Nozaki, Highly nonreciprocal spin waves excited by magnetoelastic coupling in a Ni / Si bilayer, Phys Rev Appl **13**, 034074 (2020).